\documentstyle[aps,epsf,prd,floats,cite,preprint]{revtex}

\preprint{\vbox{ \hbox{CMU-01-08} \hbox{MIT-CTP-3180} \hbox{hep-th/0109175} }}
\title{Mother Moose: Generating Extra Dimensions from Simple Groups
at Large $N$}
\author{Ira Rothstein $^a$ and Witold Skiba $^b$}
\address{ \vbox{\vskip 0.truecm} $^a$  Department of Physics,
      Carnegie Mellon University, Pittsburgh, PA 15213
 \vspace{.3cm} \\
 $^b$ Center for Theoretical Physics, Massachusetts Institute of Technology,
      Cambridge, MA 02139 \vspace{.3cm}}
\begin{document}

\maketitle
\begin{abstract}

We show that there exists a correspondence between four dimensional gauge
theories with simple groups and higher dimensional gauge theories
at large $N$. As an example, we show that a four dimensional ${\cal N}$=2
supersymmetric $SU(N)$  gauge theory, on the Higgs branch, has the same
correlators as a five dimensional $SU(N)$ gauge theory in the limit of large
$N$ provided the couplings are appropriately rescaled. We show that
our results can  be applied to the AdS/CFT correspondence to derive
correlators of five or more dimensional gauge theories from solutions
of five dimensional supergravity in the large t'Hooft coupling limit. 

\end{abstract}
\tighten
\newpage

\section{Introduction}
The idea that gauge theories in varying dimensions are intimately
related has lead to new insights to our understanding of strongly
coupled quantum field theories. One way to relate theories in differing 
dimensions is through the holographic AdS/CFT correspondence. 
Another way is to use the fact that one can trade dynamical
degrees of freedom for additional dimensions. This result  
can be seen to be a simple consequence of effective field theory.	
If we consider a $d+1$ dimensional theory and compactify one
dimension, then at distance scales large compared to the
compactification radius the theory will necessarily look $d$ dimensional.
The effects of the extra dimension are suppressed by powers of $ER$ and
can be taken into account by adding higher dimensional local operators
when integrating out Kaluza-Klein states according to a matching procedure.
Given these facts, one can see that it is possible to reverse engineer
this process and actually build an extra dimension by adding the necessary
states to the $d$ dimensional action.

Recently, this idea has been implemented within the context of ``moose''
or ``quiver'' theories. These are theories whose gauge groups are chains
of $SU(N)$'s which communicate to each other through matter in
bi-fundamental representations.
The authors of Refs.~\cite{Arkani-Hamed:2001ca,Hill:2000mu} have shown
that given a moose theory at some ultraviolet scale, symmetry breaking
will naturally lead to a five dimensional $SU(N)$ gauge theory at scales
well below the symmetry breaking scale. That is to say, there is an
approximate equivalence between Greens functions calculated in the
four dimensional theory  which include a tower of massive states,
and Greens functions calculated in a five dimensional theory. This
construction allows for sensible UV limits of theories which appear
to be higher dimensional in the IR. See also 
Refs.~\cite{Halpern:1975yj,Castro:2001ja} for related ideas.

In this paper we
extend this equivalence in the limit of large $N$. In particular
we will show that there are entire classes of $d$ dimensional
theories with simple groups which, in the large $N$ limit, have
the same correlation functions as $d+1$ dimensional theories.
It would be a straightforward exercise to extend our results to
construct $d+n$ dimensional theories. We then apply this idea to
the AdS/CFT correspondence, and show that it is possible to derive
correlators in higher dimensional ($d>4$) field theories from
classical supergravity solutions in the background of spaces
with stacks of separated D-3 branes.

The basic premise behind the generalization involves orbifolding
field theories, which we briefly review in the next section.
In Section~\ref{sec:moosing}, we outline the construction
of extra dimensions from lower dimensional field theories.
The main part of the paper is contained in Section~\ref{sec:example}
where we discuss a supersymmetric example.
The procedure of generating additional dimensions provided several
phenomenological applications~\cite{Arkani-Hamed:2001nc,Cheng:2001nh,
Csaki:2001em,Cheng:2001an,Csaki:2001qm,Cheng:2001qp,Arkani-Hamed:2001vr},
but we do not attempt to apply our work to model building.
In Section~\ref{sec:gauge-gravity}, we work out an explicit example 
where the AdS/CFT correspondence can be utilized to calculate correlators
in a five dimensional theory with sixteen supercharges.

\section{Orbifolding}
\label{sec:orbifolding}

In the large $N$ limit it is possible to relate Greens functions
for two disparate theories in a non-trivial fashion. In particular,
it has been shown that planar diagrams of two theories related by
orbifolding are identical when the gauge couplings are rescaled.
Orbifolding a field theory~\cite{Kachru:1998ys,
Bershadsky:1998mb,Kakushadze:1998tr,Bershadsky:1998cb}
entails removing a set of fields which are not invariant
under certain discrete group transformations. 

The orbifolding procedure works as follows. Consider a discrete group $G$,
with $\Gamma$ elements, embedded in an $SU(\Gamma N)$ gauge theory. For
each element of the discrete group form an $N$-fold copy of the regular
representation $\gamma^a$. With a convenient choice of basis, this
entails placing $d_i N$ copies of each irreducible representation of
$G$ with dimension $d_i$ along the diagonal. This $N$-fold copy of the
regular representation of $G$ forms a $\Gamma N \times \Gamma N$ matrix,
which is an element of $SU(\Gamma N)$. The dimensionality
works out because $\sum_i (d_i)^2=\Gamma$. The fundamental $F$ and
adjoint $A$ of the group therefore transform as
\begin{equation}
\label{transformations}
  F \rightarrow \gamma^a F, ~~~A\rightarrow \gamma^a A (\gamma^a)^\dagger.
\end{equation}
Orbifolding means that the ``daughter theory'' field content is obtained
by keeping only fields invariant under the discrete transformation in
Eq.~(\ref{transformations}). Moreover, the action is obtained from a
``mother theory''action by setting to zero all terms that contain
fields not invariant under such transformations. All terms in the action
with fields removed by the orbifolding procedure are therefore discarded.
For a more detailed description of the orbifolding procedure see
Ref.~\cite{Schmaltz:1999bg}. To get a non-trivial orbifolding,
i.e.\ one that does not lead to a daughter theory which is simply multiple
copies of the mother, it is necessary to embed the discrete symmetry in
global symmetry groups of the action, as we will see below.

It has been shown~\cite{Bershadsky:1998cb}, that the correlation
functions of the daughter theory  are identical to those of the
mother theory up to the following rescaling of the couplings
$g_m\rightarrow g_d/\sqrt{\Gamma}$. This is a non-trivial result given
that the loops of the mother theory incorporate fields which are not
present in the daughter theory.  The proof of the relation 
relies only on the fact that the leading order graphs
in $1/N$ are planar. The correspondence is thus strictly true
perturbatively. However, there is considerable evidence, especially
in supersymmetric theories, that it works at the non-perturbative
level as well \cite{Schmaltz:1999bg,Erlich:1998gb,Strassler:2001fs}.
Here we will utilize the orbifolding procedure to generate moose
theories which, upon symmetry breaking, generate higher dimensional
theories.

\section{Moosing}
\label{sec:moosing}

Consider a theory of $\Gamma$ $SU(N)$ gauge groups which are
linked together by matter in the bi-fundamental representation.
The theory may be depicted as in Figure 1, with each circle
representing a gauge group and each line representing a matter
multiplet~\cite{Georgi:1986hf}.
An arrow pointing into (out of) a gauge group corresponds
to a matter multiplet transforming in the (anti)fundamental 
representation of that particular gauge group. The gauge symmetry
is broken down to the diagonal subgroup at scale $v$, when the
bi-fundamental fields get expectation values. The bi-fundamental
field may be either fundamental scalars or fermionic bound states.
We could represent them as linear or non-linear sigma models. In the
following we will use the linear realization, so the Lagrangian
of the moose configuration is 
\begin{equation}
\label{latticeaction}
  {\mathcal L}=\sum_{i=1}^\Gamma \, Tr\left(-\frac{1}{4g^2}F^{\mu \nu i} 
  F_{\mu \nu}^i+(D^{\mu}U_{i})^{\dagger}D_{\mu}U_{i}\right)\, +\, \ldots,
\end{equation}
where we have omitted the potential for the matter fields which is necessary
for giving vevs to those fields. The covariant derivatives are defined as
$D_\mu U_i=\partial_\mu U_i -i A^i_\mu U_i+ i U_i A_\mu^{i+1}$.
When all $U$'s obtain diagonal vevs, $U_i=v {\mathbf 1} + \hat{U}_i$,
this action is exactly the latticized version of the action of a compact
$D+1$ dimensional $SU(N)$ gauge theory of radius $R=\Gamma a$ and
lattice spacings $a=\frac{1}{gv}$. The $U$'s play the role of the link
variable in the latticized new dimension.

\begin{figure}
\epsfxsize=5cm
\hfil\epsfbox{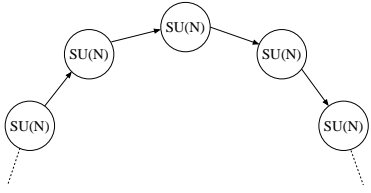}\hfill
\caption{Typical moose diagram
\label{fig:pair}}
\end{figure}

At energy scales in the range $gv/\Gamma<< E << gv$ the theory
looks like a $D+1$ dimensional continuum gauge theory. This may be
thought of as essentially the reverse process of Kaluza-Klein reduction.
If we compactify a $D+1$ dimensional theory on a flat background then
the $D$ dimensional action is just comprised of zero modes for the
fields plus a tower of massive KK states with masses
\begin{equation}
  m_n=\frac{2\pi n}{R}.
\end{equation}
The moose is simply a construct which naturally leads to a
tower of states, which in the above energy range, and for
large enough $\Gamma$, approximately reproduce the spectrum of KK modes. 
Indeed, if we diagonalize the mass matrix for the gauge bosons
in Eq.(\ref{latticeaction}), we find~\cite{Arkani-Hamed:2001ca,Hill:2000mu}
\begin{equation}
  M_n^2=4g^2v^2 sin^2\left( \frac{\pi n}{\Gamma}\right),
\end{equation}
where $-\Gamma/2 < n \leq \Gamma/2$. For $n<<\Gamma$ this approximates 
the above spectrum for the KK tower.
Notice that the gauge coupling of the diagonal subgroup is given
by 
\begin{equation}
  \frac{1}{g^2_D}=\frac{R}{g_{D+1}^2}.
\end{equation}

It is simple to see how the extra-dimensional phase space is
generated once the sum over the KK tower is performed.
Consider two particle phase space for the scattering of a zero mode:
\begin{equation}
  2PS\equiv\sum_n \int \frac{d^dk_1^n}{(2\pi)^{d-1}E_1^n} \frac{d^dk_2^n}
  {(2\pi)^{d-1}E_2^n} \delta^{(d)}(P-k_1^n-k_2^n),
\end{equation} 
where $E^n=\sqrt{k^2+m_n^2}$ is the energy of the $n$-th KK mode.
In flat space K-K number is conserved, thus both final states
have the same masses. In the limit where $\Gamma>>1$ we may replace
the sum over KK modes by an integral such that
\begin{eqnarray}
  2PS &\equiv& \int dm dm^\prime 
       \frac{{d^dk_1}}{(2\pi)^{d-1}E_1(m)} \frac{{d^dk_2}}
       {(2\pi)^{d-1}E_2(m)} \delta^{(d)}(P-k_1-k_2)\delta(m-m^\prime)
         \nonumber \\
  &\propto& \int
   \frac{{d^{d+1}k_1}}{(2\pi)^{d}E_1} \frac{{d^{d+1}k_2}}
   {(2\pi)^{d}E_2} \delta^{(d+1)}(P-k_1-k_2).
\end{eqnarray}
The second delta function in the first line  enforces conservation
of momentum in the extra dimension. This relation had to be true
if the full exact KK tower were included, and is approximately
true, up to power corrections, in the moose construction.

\section{Extra Dimensions from Simple Groups}
\label{sec:example}

We would now like to combine these two techniques in such a way
that we may generate an extra dimension from theories which
are simpler than mooses. In particular, we would like to show
that at large $N$ there are simple groups which, when Higgsed, 
generate extra dimensions. Here we will consider one simple
supersymmetric example in four dimensions. In the supersymmetric
case the role of the bi-fundamental field will be played by the scalar
component of a chiral multiplet, but it is also possible to generate
non-supersymmetric examples.

One might be concerned that the spontaneous symmetry breaking
may destroy the simple relation between planar diagrams. 
It is important that the choice of the vev does not break the orbifold
symmetry. This happens automatically if the fields which obtain
vevs are those shared by the mother and daughter theories. Such
fields transform trivially under the orbifold symmetry. The
correspondence between orbifolded theories holds diagram by
diagram. Given corresponding diagrams in unbroken phases, 
we can compare such diagrams in spontaneously broken phases.
For the correspondence to go through in the broken phases
of the theory we must impose the constraint that $g v$ be the
same in the two theories. That is, $v_m=\sqrt{\Gamma} v_d$.
This condition arises from that fact that, when the gauge fields
are canonically normalized, an $n$ particle interaction vertex
must scale as $g^{n-2}$ in order for the orbifold correspondence
to hold\cite{Schmaltz:1999bg}.

We begin with an ${\mathcal N}=1$ supersymmetric $SU(\Gamma N)$ gauge
theory with a chiral superfield $A$ in the adjoint representation.
In principle we could add a superpotential for the field $A$, which can be
made consistent with planarity in the large $N$ limit, but for simplicity
we will set the superpotential to zero. Therefore this theory is equivalent
to pure ${\cal N}=2$ SUSY Yang-Mills theory. In component form this theory
has Lagrangian
\begin{equation}
\label{action}
  {\mathcal L}=\frac{1}{g^2}{\rm Tr}\left( -\frac{1}{4}F^{\mu \nu}F_{\mu \nu}
  -i\bar{\lambda} D \!\!\!\! \slash
  \lambda-i\bar{\psi}  D\!\!\!\! \slash \psi+
  \mid\!\! D_\mu \phi\!\! \mid^2-\frac{1}{2}\left[\phi^\dagger,\phi\right]^2
  -i\sqrt{2}[\lambda,\psi]\phi^\dagger
  -i\sqrt{2}[\bar{\lambda},\bar{\psi}]
\phi\right),
\end{equation}
where the gauge field and $\lambda$ are components of the vector multiplet,
while $\psi$ and $\phi$ are components of the chiral multiplet $A$.

This theory has an extensive moduli space parameterized by
the set of holomorphic gauge invariant polynomials. 
The independent gauge invariants are powers of the adjoint:
${\rm Tr}( A^i)$, $i=2,\ldots,\Gamma N$. On the Coulomb branch the
low energy effective action can be solved for up to higher
derivative terms \cite{Seiberg:1994rs}. Here will be not be
interested in the Coulomb branch but in the Higgs branch instead.

We consider orbifolding an $SU(\Gamma N)$ by the discrete subgroup 
$Z_\Gamma$. The gauge indices of all the multiplets will transform
according to Eq.~(\ref{transformations}) under this orbifolding.
In addition, we will embed the $Z_\Gamma$ discrete symmetry
into a $U(1)_A$ global symmetry of the chiral superfield. The $U(1)_A$
rotates the adjoint matter field by a phase, so it is an ``adjoint number''
symmetry. This $U(1)$ symmetry is anomalous with a $Z_{2 \Gamma N}$
subgroup preserved by instantons. We embed the ``orbifold''
symmetry $Z_\Gamma$ into the non-anomalous subgroup of $U(1)_A$.

Under orbifolding the $SU(\Gamma N)$ gauge group breaks up into the product
group  $SU(N)^\Gamma$. There are $\Gamma-1$ additional  $U(1)$ gauge
symmetries preserved by the orbifolding. With our choice of
matter content these $U(1)$ symmetries will be anomalous, but do not
contribute in the large N limit. The matter fields which are kept in
the daughter theory are those which are invariant under the transformation
\begin{equation}
\label{phase}
  (\psi,\phi)\rightarrow r^a \gamma_a (\psi,\phi) (\gamma_a)^\dagger,
\end{equation}
where $r^a$ is a pure phase, and has the effect of pushing the invariant
fields to the right of the diagonal. Schematically, the orbifolding gives
\begin{equation}
\label{fields}
A \rightarrow \left( \begin{array}{ccccc}
                     & A_1 & &\ldots &  \\
                     & & A_2 & \ldots & \\
                    \vdots & \vdots& \vdots & \ddots & \vdots \\
                     & & & \ldots & A_{\Gamma-1} \\
                     A_\Gamma & & & \ldots &
                     \end{array} \right),
\end{equation}
where $A_i$ are the daughter theory bi-fundamentals transforming under
neighbor $SU(N)$ groups. In performing the orbifolding we have
reduced the number of supersymmetries to four (we will regain
supersymmetry in the low energy limit where the theory looks five dimensional).
This is precisely the theory illustrated by the
moose diagram  shown in Figure 1. The Seiberg-Witten curves for the Coulomb
branch of this theory were derived in Ref.~\cite{Csaki:1997zg}. 

For our purposes we will choose a point on the moduli space corresponding
to a partial Higgsing of the gauge group. Our choice of a points on the
moduli space is such that, in the daughter theory, the bi-fundamental
scalars each get equal expectation values. The corresponding vev in the
mother theory is of the form
\begin{equation}
\label{vevs}
\langle A \rangle = v \left( \begin{array}{ccccc}
                     & {\mathbf 1} & &\ldots &  \\
                     & & {\mathbf 1} & \ldots & \\
                    \vdots & \vdots& \vdots & \ddots & \vdots \\
                     & & & \ldots & {\mathbf 1} \\
                     {\mathbf 1} & & & \ldots &
                     \end{array} \right),
\end{equation}
where each ${\mathbf 1}$ represents an $N\times N$ identity matrix.
This choice of vevs is part of the moduli space of both mother and
daughter theories. In the daughter theory all vevs are aligned and
proportional to the identity, so they are D-flat. They correspond to baryonic
flat directions. In the mother theory it is easy to
check that $\langle A \rangle \, \langle A^\dagger \rangle - 
\langle A^\dagger \rangle \, \langle A \rangle =0$, therefore
all D terms vanish~\cite{Affleck:1985xz}.

Breaking the $SU(\Gamma N)$ gauge symmetry with the vev of Eq.~(\ref{vevs})
leaves the mother theory with an $SU(N)^\Gamma$ gauge symmetry.
It is easy to see the breaking pattern after diagonalizing the vev in the
mother theory
\begin{equation}
\label{diagonalvevs}
\langle A \rangle = v \left( \begin{array}{ccccc}
                    {\mathbf 1} &  & &\ldots &  \\
                     & \omega\, {\mathbf 1} &  & \ldots & \\
                     & & \omega^2 \,{\mathbf 1} & \ldots & \\
                    \vdots & \vdots & \vdots & \ddots & \vdots \\
                    & & & \ldots & \omega^{\Gamma-1}\, {\mathbf 1}
                     \end{array} \right),
\end{equation}
where $\omega^\Gamma=1$. In this new basis it is easy to read of the 
spectrum of the mother theory. The diagonal $N \times N$ blocks
correspond to unbroken $SU(N)^\Gamma$, each with a massless adjoint
superfield. In addition, there are $\Gamma-1$ massless $U(1)$ gauge bosons
as well as $\Gamma-1$ massless chiral superfields which are singlets
under all the unbroken gauge groups. These fields do not contribute at the 
leading order in $N$. The off-diagonal $N \times N$ blocks
correspond to massive vector superfields. The masses of the fields  
in different blocks are
\begin{equation}
\label{mothermasses}
M^2 = 4 g^2_m v^2_m \left( \begin{array}{ccccc}
  0 & \sin^2(\pi/\Gamma) & \sin^2(2 \pi/\Gamma)&\ldots & \sin^2(\pi/\Gamma) \\
  \sin^2(\pi/\Gamma) & 0 & \sin^2(\pi/\Gamma) & \ldots & \sin^2(2\pi/\Gamma)\\
  \sin^2(2\pi/\Gamma) & \sin^2(\pi/\Gamma) & 0 & \ldots & \sin^2(3\pi/\Gamma)\\
  \vdots & \vdots & \vdots & \ddots & \vdots \\
  \sin^2(\pi/\Gamma) & \sin^2(2\pi/\Gamma) & \sin^2(3\pi/\Gamma) & \ldots & 0
                     \end{array} \right).
\end{equation}

The vev of Eq.~(\ref{vevs}) breaks the daughter theory to a diagonal $SU(N)$
with a gauge coupling $g_{diag}^2=g_d^2/\Gamma$. In addition there is a
massless chiral superfield transforming in the adjoint of the diagonal $SU(N)$.
The massless gauge bosons are accompanied by a tower of massive
gauge bosons with masses
\begin{equation}
\label{daughtermasses}
  m^2_k=4g^2_d v^2_d \sin^2(\frac{\pi k}{\Gamma}).
\end{equation}
Note that if we want to keep the masses fixed as we take the t'Hooft limit
$N \rightarrow \infty$ with $g^2 N$ fixed, we need to simultaneously
rescale the vevs and keep $v^2/N$ unchanged. From now on, we will assume
that $g v$ is held fixed in the large $N$ limit. There are also $\Gamma$
massless singlet chiral superfields, which are the parts of the link
fields not eaten by the super-Higgs mechanism when breaking to the
diagonal subgroup. 

Comparing Eqs.~(\ref{mothermasses}) and (\ref{daughtermasses}) we obtain the
same mass spectrum if $v_m=v_d \sqrt{\Gamma}$, in concordance with
our previous statement of the equality of $gv$ in the two theories.
In the interesting limit of $k<<\Gamma$ we get 
$m_k\simeq k\, (2 \pi g v)/ \Gamma$.
If we define the lattice spacing as $a^{-1}=gv$ and the compactification
radius as $R=\Gamma a$, then we have correctly reproduced the KK tower
for a compactified flat extra dimension. The pattern of orbifolding and
symmetry breaking is shown in Figure 2.

\begin{figure}
\epsfxsize=5cm
\hfil\epsfbox{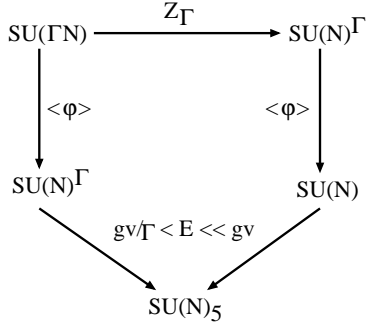}\hfill
\caption{The pattern of orbifolding and symmetry breaking.
The mother theory is on the left and the daughter on the right.}
\end{figure}

In order to compare the correlators of the mother and daughter
we need to identify the corresponding fields in the two theories.
It is only for these shared fields  that we can expect the
equality of correlators. In the basis defined by Eq.~(\ref{fields})
the correspondence between the fields is most transparent.
The corresponding mother's and daughter's fields reside at the same
positions. Vectors live on the diagonal blocks and scalars live on blocks
above the diagonal. Since in the mother we changed the basis by
diagonalizing vevs in Eq.~(\ref{diagonalvevs}), we need to
perform the same transformation on all the fields.

Let us start with the massless gauge multiplet
of the daughter theory. This multiplet, after the breaking of
the $SU(N)^\Gamma\rightarrow SU(N)_{diag}$ symmetry, is a linear
combination of all $SU(N)^\Gamma$ fields with the same coefficient
for each field. The corresponding field in the mother theory is the same
linear combination of $SU(N)^\Gamma$ fields on the diagonal.
For this special linear combination there is no change induced in  going
from basis defined by Eq.~(\ref{vevs})
to the one defined by Eq.~(\ref{diagonalvevs}). 
The properly normalized linear combination of fields is
\begin{equation}
  A^a_{diag}=(A^a_1+A^a_2+\ldots+A^a_{\Gamma})/\sqrt{\Gamma}.
\end{equation}
Therefore, the coupling of the diagonal subgroup is $g/\sqrt{\Gamma}$.

As far as the massive multiplets are concerned
each $N\times N$ block of the mother theory transforms as
an adjoint under the diagonal subgroup since bi-fundamentals of
$SU(N)^\Gamma$ transform as adjoints under the diagonal subgroup. With
this identification we begin to see why the correlators are identical up
to a rescaling. At each mass level of the mother theory there are $\Gamma$
adjoints of the diagonal subgroup. However, each of the adjoints couples
with the reduced coupling $g/\sqrt{\Gamma}$, so the two compensate.
The rescaling of the coupling then compensates for the factor
of $\sqrt{\Gamma}$ in the daughter theory coupling.
In the large $N$ limit, additional massless fields whose number does
not grow with $N$ become irrelevant. Since the spectra of the mother
and daughter differ only by $O(\Gamma)$ fields, these theories
have identical behavior for large $N$.

The massive vector multiplets in the daughter, again, correspond
in the mother theory to linear combinations of fields. 
At mass level $k$ these are 
\begin{equation}
  A^a_k=(A^a_1+\omega^k A^a_2+\omega^{2 k}\ldots+
  \omega^{k(\Gamma-1)} A^a_{\Gamma})/\sqrt{\Gamma}
\end{equation}
in the daughter  theory, as well as in the mother theory before
diagonalizing the vev. When changing to the basis defined by
Eq.~(\ref{diagonalvevs}) we need to appropriately transform this linear
combination. It turns out that in the new basis the linear combinations
no longer involves fields residing in the diagonal blocks, instead it
involves all fields from blocks parallel to the diagonal. The linear
combination includes vector fields from each $N\times N$ block with a
given mass as displayed in Eq.~(\ref{mothermasses}). For the chiral
superfields the linear combinations have different coefficients than those
for the vector fields with the same mass. However, in the basis
with the diagonal vev the linear combination of scalars also involves
all fields in the appropriate blocks parallel to the diagonal.
This is, of course, expected since the vector and chiral multiplets
must combine into irreducible massive vector multiplets.

We can single out the diagonal subgroup of the mother theory,
for which correlators look five dimensional, in several ways.
First, we can impose a discrete symmetry that identifies
$\Gamma$ $SU(N)$ factors. Alternatively, in the mother theory
we can add a small diagonal vev with magnitude $v'$, which breaks
$SU(N)^\Gamma$ to the diagonal. When $v'\ll v$ the splittings due to
$v'$ are negligible compared to the K-K splittings induced by $v$.
Thus, we get $\Gamma$ nearly degenerate adjoints of the diagonal
subgroup at each mass level $g v |\sin(\pi k/\Gamma)|$.

As we already mentioned, one can extend this construction to generate
more dimensions by orbifolding a simple group. For example, an
$SU(\Gamma^2 N)$ theory with two adjoint chiral superfields can be
orbifolded by a $Z_\Gamma \times Z_\Gamma$ discrete symmetry. We assign
the transformation properties of the two chiral fields similar to
Eq.~(\ref{phase}), such that each of them picks a phase under one of
$Z_\Gamma$ factors. The daughter theory resembles a two dimensional lattice,
so with appropriate vevs both mother and daughter generate two extra
dimensions. 

For the daughter theory to look five dimensional we must restrict
ourselves to the regime
\begin{equation}
\label{energyrange}
  g_d v_d/\Gamma<<E<<g_d v_d.
\end{equation}
Now we must ask, what does the mother theory look like at
these energy scales? We would expect that at some scale, $\Lambda_m$,
the mother theory will get strong, and will most likely undergo a
phase transition. However, the orbifold correspondence is blind
to such effects. That is, the correspondence tells us that given the
Greens function in the mother theory we can get the Greens
functions in the daughter by simply making the replacement
$g_m \rightarrow g_d /  \sqrt{\Gamma}$, or vice versa. This statement is
independent of scales. What does this mean in terms of our correspondence?
It would seem that we are allowed to take the mother theory to be as
weakly coupled or as strongly coupled as we wish and that in either case
we can still equate its Greens functions with those of the 5-d theory.
However,  in order for the daughter theory to
look five dimensional, we can only study energy scales with 
$E>> g v_d/\Gamma$. We are free to take $\Gamma$ to be as large
as we wish, thus studying the theories at arbitrarily large
distances. However, the four dimensional mother theory will get its
coupling rescaled by $1/ \sqrt{\Gamma}$, so it will remain weak in
the regime where the daughter looks five dimensional.
This had better be the case since the five dimensional theory is
weakly interacting in the IR. Indeed, the gauge interaction 
in five dimensions is irrelevant and the theory is IR free.
In the context of the four dimensional theory this can be seen
from the fact that as we lower the energy scale more and more
KK modes are decoupling and thus the interaction is getting weaker.
So the effective coupling will scale with $E\Gamma/(g v)$,
as one would expect for a compactified five dimensional theory.

\section{Gauge Gravity Correspondence in Five Dimensions}
\label{sec:gauge-gravity}

Suppose we start with an $SU(\Gamma N)$ mother theory which is ${\cal N}=4$
supersymmetric. By embedding the discrete orbifold group inside the $SU(4)$
R symmetry we may relate the mother theory correlation functions
to those of an ${\cal N}=2$ (or ${\cal N}=1,0$) supersymmetric daughter
theory. Given that the mother theory is finite and has vanishing
beta function, the daughter theory must, in the large $N$ limit, be finite
as well. In this section, we will assume that the mother is orbifolded 
to an ${\cal N}=2$ daughter. We can achieve that by embedding $Z_\Gamma$ in
the R symmetry with matrix $diag(\omega,\omega^*,1,1)$ acting on the
fundamental representation of $SU(4)_R$. The matter content of the daughter 
is composed of $\Gamma$ vector multiplets and $\Gamma$ bi-fundamental
hyper-multiplets. In terms of the ${\cal N}=2$ multiplets, the theory has
identical structure as the moose depicted in Figure 1 (with links now
unoriented). Moving away from the origin of moduli space onto the Higgs
branch we may moose the theory in the same way as in the example discussed
above. The resulting five dimensional theory has 16 supercharges.
The mother starts out as a superconformal theory, however going out
on the Higgs branch maintains 16 supercharges. The daughter has eight
supercharges and additional accidental supersymmetry at low energies.
 
In the following, we will outline how the AdS/CFT correspondence 
might be used to calculate correlators of five dimensional field theory.
The AdS/CFT correspondence is commonly used for field theories with
large t'Hooft coupling. One might worry that the discussion of the
spontaneous symmetry breaking is only applicable at weak coupling.
However, in ${\cal N}=4$ theory all masses are related to the central
charges of the supersymmetric algebra and are BPS saturated.
Thus, the tree-level results are not renormalized and one obtains
the same KK tower for any of the value of the gauge coupling.

The deformation onto the Higgs branch yields a non-trivial RG
flow to an IR fixed point, which is reached once we integrate
out all the KK modes. This scenario has an interesting interpretation
in the AdS/CFT correspondence. Recall that the $D3$ brane metric in
type IIB string theory is given by
\begin{equation}
  ds^2=H^{-1/2} \eta_{\mu \nu}dx^\mu dx^\nu+H^{1/2}\delta_{ij}dy^idy^j,
\end{equation}
with the $x^\mu$ coordinates spanning the world-volume of the D brane
and $y^i$ span ${\bf R}^6$. H is given by
\begin{equation}
  H=h+\frac{R^4}{r^4},
\end{equation} 
where $R^4\equiv 4 \pi g_s \alpha^\prime N$, 
$r^2=\delta_{ij}y^iy^j$, and $h=1$. Near $r\rightarrow 0$, one
approaches a non-singular horizon, and the space is locally $AdS_5\times S^5$.
Whereas, in the limit $r\rightarrow \infty$ approaches 10 dimensional
flat space. 

The case of $h=0$, which is $AdS_5 \times S_5$ for all $r$,
is holographically described by the ${\cal N}=4$ super-conformal field theory,
whereas the $h=1$ case retains 16 supercharges but is no longer conformal.
It flows in the IR ($r\rightarrow 0$) to the $h=0$ case. The $h=1$ case
has been conjectured to be dual to a super-conformal theory perturbed by a
dimension eight operator~\cite{Gubser:1999iu,Intriligator:2000ai}.
It turns out that this dimension eight operator is the most relevant
operator induced by integrating out massive states in a Higgsed
theory~\cite{Intriligator:2000ai,Costa:2000gk}.

We wish to Higgs the $SU(\Gamma N)$ mother theory, in the fashion described
in the previous section. This may be accomplished by separating the stack of
$\Gamma N$ D3 branes, leaving the geometry
\begin{equation}
  H(y)=h+4 \pi g_s \alpha^\prime \sum_{i=1}^k \frac{N_i}
  {\mid\vec{y}-\vec{a_i}\mid^4},
\end{equation}
where $N_i$ are the number of D3 branes places at position $\vec{a}_i$
in the transverse space, and $\sum_{i=1}^k N_i=\Gamma N$. 
This solution breaks the $SU(4)$ R symmetry but
preserves the 16 supercharges. The $\vec{a}_i$ correspond to the
diagonal vevs of the six real adjoint scalars. The above brane
configuration is only applicable at small values of the string coupling,
so it is not apparent how to describe weakly coupled field theories this way. 

Let us now consider the symmetry
breaking pattern required for the moose constructions described in
Eq.~(\ref{diagonalvevs}). In this case there are only two linearly
independent vevs of the real scalar adjoints. One of these fields has
vevs proportional to the real parts of $\omega^k$, while the other
to the imaginary parts, where $k=0,\ldots,\Gamma-1$. This means that we
can place all branes in just two transverse directions. In field theory
this reflects the breaking of the $SU(4)\sim SO(6)$ R symmetry to
$SO(4)$. We pick out two of the transverse directions, say $y^5$ and
$y^6$ and place $\Gamma$ groups of $N$ branes with coordinates
\begin{eqnarray}
  a^5 & \propto & v \, \{ 1,\cos(2 \pi/\Gamma), \cos(4 \pi/\Gamma),
                      \ldots, \cos(2 \pi(\Gamma -1)/\Gamma) \}, \nonumber \\
  a^6 & \propto & v \, \{ 0,\sin(2 \pi/\Gamma), \sin(4\pi/\Gamma),
                      \ldots,\sin(2 \pi(\Gamma -1)/\Gamma) \}.
\end{eqnarray}
As might have been anticipated, the branes are arranged along the perimeter
of a circle. Similar dualities of five dimensional field theories
can be obtained from the orbifolds of the $AdS \times S^5$ space,
which were described in Refs.~\cite{Kachru:1998ys,Bershadsky:1998mb,
Kakushadze:1998tr,Bershadsky:1998cb}.

The field theory dual of this brane construction corresponds, by orbifolding
at large $N$, to a five-dimensional theory.  While the four dimensional
coupling does not run, the effective coupling, which accounts for the fact
that as we flow into the IR more and more KK modes are decoupling, becomes
weaker. This is consistent with the five dimensional interpretation of the
theory, where the coupling is an irrelevant operator at the classical level.

The five-dimensional field theory interpretation only applies for energies
in the range $gv/\Gamma<<E<<gv$. For the supergravity approximation in
the AdS/CFT correspondence to be trustworthy, the AdS curvature scale must
be  small compared to the Planck scale. This corresponds to taking
$g_m \sqrt{N \Gamma}>>1$. The daughter's coupling is
$g_m \sqrt{\Gamma}$, but for the five dimensional theory
the relevant coupling is that of the diagonal subgroup. Therefore,
the effective coupling in the five-dimensional theory is
\begin{equation}
  (ER) \frac{g^2_dN}{16\pi^2}=\frac{g^2_mN}{16 \pi^2}
    \frac{E \Gamma}{g_m v_m}.
\end{equation}
The five-dimensional interpretation makes sense only for energies
below the cut-off, that is in the region where the loop expansion
parameter in the effective theory is less than one. Thus, the
energy range must be restricted to
\begin{equation}
  g_m v_m/\Gamma<<E < g_m v_m \frac{16 \pi^2}{g^2_m N \Gamma},
\end{equation} 
where the lower bound is equal to the mass of the lightest KK
state. If $g_m \sqrt{N \Gamma} < 4 \pi$ then our AdS/CFT dual can
be interpreted as a five-dimensional field theory in the full
energy range $gv/\Gamma<<E<<gv$ accessible to moose constructions.  
If the t'Hooft coupling is large $g_m \sqrt{N \Gamma} > 4 \pi$ then
only part of that energy range remains below the cut-off. Of course,
we cannot take the t'Hooft coupling to be too small or string corrections
will become important. It is not difficult to generalize this
construction to allow for a correspondence between $AdS_5$ and gauge
theories of dimension higher than five. One needs to simply choose the
appropriate $D$ brane configurations in the bulk. It would be
interesting to search for classical solutions in these non-trivial
backgrounds to test the correspondence with higher dimensional field
theories.

{\bf Note added:}
After completing this work we  were made aware of
Refs.~\cite{Sfetsos:1999xd,Sfetsos:2001qb} which have
considerable overlap with Section~\ref{sec:gauge-gravity}.

\acknowledgements
       
The authors indebted  the Aspen Center for Physics, where this work
was initiated, for its hospitality. We would like to thank Dan Freedman,
Walter Goldberger, Ken Intriligator, Andreas Karch, and Martin Schmaltz
for discussions. The work of I.R. is supported in part by the Department
of Energy under grant DOE-ER-40682-143, while that of W.S. by grant
DE-FC02-94ER40818.


 \end{document}